\documentstyle[epsf]{mn}

\newif\ifAMStwofonts
\def\simlt{\lower.5ex\hbox{$\; \buildrel < \over \sim \;$}}
\def\simgt{\lower.5ex\hbox{$\; \buildrel > \over \sim \;$}}

\def\be{\begin{equation}}
\def\ee{\end{equation}}

\def\bee{\begin{eqnarray}}
\def\eee{\end{eqnarray}}

\ifoldfss
  \ifCUPmtlplainloaded \else
    \NewTextAlphabet{textbfit} {cmbxti10} {}
    \NewTextAlphabet{textbfss} {cmssbx10} {}
    \NewMathAlphabet{mathbfit} {cmbxti10} {} % for math mode
    \NewMathAlphabet{mathbfss} {cmssbx10} {} %  "   "    "
  \fi
  \ifAMStwofonts
    \ifCUPmtlplainloaded \else
      \NewSymbolFont{upmath} {eurm10}
      \NewSymbolFont{AMSa} {msam10}
      \NewMathSymbol{\upi}     {0}{upmath}{19}
      \NewMathSymbol{\umu}     {0}{upmath}{16}
      \NewMathSymbol{\upartial}{0}{upmath}{40}
      \NewMathSymbol{\leqslant}{3}{AMSa}{36}
      \NewMathSymbol{\geqslant}{3}{AMSa}{3E}

    \fi
  \fi
\fi % End of OFSS

\ifnfssone
  \newmathalphabet{\mathit}
  \addtoversion{normal}{\mathit}{cmr}{m}{it}
  \addtoversion{bold}{\mathit}{cmr}{bx}{it}
  \newmathalphabet{\mathbfit} % math mode version of \textbfit{..}
  \addtoversion{normal}{\mathbfit}{cmr}{bx}{it}
  \addtoversion{bold}{\mathbfit}{cmr}{bx}{it}
  \newmathalphabet{\mathbfss} % math mode version of \textbfss{..}
  \addtoversion{normal}{\mathbfss}{cmss}{bx}{n}
  \addtoversion{bold}{\mathbfss}{cmss}{bx}{n}
  \ifAMStwofonts
    \ifCUPmtlplainloaded \else
      \UseAMStwoboldmath
      \makeatletter
      \new@mathgroup\upmath@group
      \define@mathgroup\mv@normal\upmath@group{eur}{m}{n}
      \define@mathgroup\mv@bold\upmath@group{eur}{b}{n}
      \edef\UPM{\hexnumber\upmath@group}
      \new@mathgroup\amsa@group
      \define@mathgroup\mv@normal\amsa@group{msa}{m}{n}
      \define@mathgroup\mv@bold\amsa@group{msa}{m}{n}
      \edef\AMSa{\hexnumber\amsa@group}
      \makeatother
      \mathchardef\upi="0\UPM19
      \mathchardef\umu="0\UPM16
      \mathchardef\upartial="0\UPM40
      \mathchardef\leqslant="3\AMSa36
      \mathchardef\geqslant="3\AMSa3E
    \fi
  \fi
\fi % End of NFSS release 1

\ifnfsstwo
  \DeclareMathAlphabet{\mathbfit}{OT1}{cmr}{bx}{it}
  \SetMathAlphabet\mathbfit{bold}{OT1}{cmr}{bx}{it}
  \DeclareMathAlphabet{\mathbfss}{OT1}{cmss}{bx}{n}
  \SetMathAlphabet\mathbfss{bold}{OT1}{cmss}{bx}{n}
  \ifAMStwofonts
    \ifCUPmtlplainloaded \else
      \DeclareSymbolFont{UPM}{U}{eur}{m}{n}
      \SetSymbolFont{UPM}{bold}{U}{eur}{b}{n}
      \DeclareSymbolFont{AMSa}{U}{msa}{m}{n}
      \DeclareMathSymbol{\upi}{0}{UPM}{"19}
      \DeclareMathSymbol{\umu}{0}{UPM}{"16}
      \DeclareMathSymbol{\upartial}{0}{UPM}{"40}
      \DeclareMathSymbol{\leqslant}{3}{AMSa}{"36}
      \DeclareMathSymbol{\geqslant}{3}{AMSa}{"3E}
    \fi
  \fi
\fi % End of NFSS release 2

\ifCUPmtlplainloaded \else
  \ifAMStwofonts \else % If no AMS fonts
    \def\upi{\pi}
    \def\umu{\mu}
    \def\upartial{\partial}
  \fi
\fi

\title[Temperature profiles and SZ]
{Cluster temperature profiles and 
Sunyaev-Zeldovich observations}
\author[Steen H. Hansen]
{Steen H. Hansen\\
University of Zurich, Winterthurerstrasse 190,
CH-8057 Zurich, Switzerland }
\date{Draft version \today}
\pagerange{\pageref{firstpage}--\pageref{lastpage}}
\pubyear{2002}

\begin{document}

\maketitle

\label{firstpage}

\begin{abstract}
Galaxy clusters are not iso-thermal, and the radial temperature
dependence will affect the cluster parameters derived through the
observation of the Sunyaev-Zeldovich (SZ) effect. We show that the
derived peculiar velocity will be systematically shifted by
$10-20\%$. For future all-sky surveys one cannot rely on the
observationally expensive X-ray observations to remove this systematic
error, but one should instead reach for sufficient angular resolution
to perform a deprojection in the SZ spectra. The Compton weighted
electron temperature is accurately derived through SZ observations.
\end{abstract}


%PACS number(s): 98.65.Cw, 98.65.Hb

%%%%%%%%%%%%%%%%%%%%%%%%%%%%%%%%%%%%%%%%%%%%%%%%%%%%%%%%%%%%%%%%%%%%%%

\section{Introduction}
Galaxy clusters have been known and studied for many years, and the
radial dependence of cluster temperatures is becoming a testing ground
for models of structure formation and for our understanding of gas
dynamics. Galaxy clusters are not iso-thermal.  Instead, the emerging
temperature profile is one where the temperature is approximately flat
or increases from the centre to some characteristic radius, and then
decreases again for larger radii.

The central temperature 
decrement has been much discussed and the possibility of
cooling flows has been explained in excellent
reviews~\cite{fabian94,donahue03}.  Many clusters are well fit with a
power law $ T \sim r^\tau$ in the very central region~\cite{voigt03}, with
a slope, $\tau$, between $0.15$ and $0.45$.  
Numerical simulations are only now
beginning to see this central decrement (see Motl et 
al.~\shortcite{motl03} for references).

The outer temperature decrease is well established both
observationally~\cite{markevitch98,degrandi02,kaastra03,arnaud} and
numerically (see Lin et al.~\shortcite{lin03} and references therein).  
The simulations have even started
to provide estimates of the outer temperature decrement which are well
fit with power laws in fair agreement with observations.

Such non-trivial temperature profiles also affect derived
cosmological parameters like the Hubble parameter~\cite{battistelli,lin03}.

On the other hand, observations of the Sunyaev-Zeldovich (SZ) effect
are becoming increasingly accurate~\cite{laroque02,coma,battistelli},
however, most analyses of SZ observational data are made under the
simplifying assumption of iso-thermality. 
Several groups have considered the ability of future SZ observations to 
extract cluster parameters~\cite{knox,AHL}, but always under the assumption 
of iso-thermality.
We therefore set out to
study the effect on the SZ derived parameters of non-trivial cluster
temperature profiles.

In section~\ref{sec:sz} we show that the relevant quantities for SZ
observations are Compton-averaged, in particular we emphasize the
difference between the temperatures derived through X-ray and SZ
observation for clusters which are not iso-thermal.  In
section~\ref{sec:cluster} we use observed and simulated 
cluster profiles to show
that the systematic shift in the derived peculiar velocity, $v_p$, is
of the order $10-20\%$. Finally we comment on the possibility of
deprojection of SZ observed spectra in section~\ref{sec:onion}.

\section{The SZ effect}
\label{sec:sz}

The SZ effect is dominated by the Compton parameter
\be
y = \sigma _T \int  n_e \frac{kT_e}{m_e c^2 }  \, dl \, ,
\ee
where $\sigma_T$ is the Thomson cross section, 
$n_e,T_e$  and $m_e$ are number density, temperature
and mass
of the electrons, and the integral is along the line of sight.
If we extract the radial dependence of the 
parameters~\cite{bha91}
\begin{eqnarray}
n_e(r) &=& n_e^0 \, f_{n_e}(r) \, , \nonumber \\
T_e(r) &=& T_e^0 \, f_{T_e}(r) \, , \nonumber 
\end{eqnarray}
then $y$ can be written as $y=\kappa_y \int f_{n_e} f_{T_e} \, dl$,
where the constant is $\kappa_y = \sigma_T n_e^0 k T_e^0/(m_e c^2)$.
It will be convenient to introduce $f_y = f_{n_e} f_{T_e}$, in which
case we have
\be
y = \kappa_y \int f_y \, dl \,.
\ee
Similarly, the optical depth is 
\be
\tau = \kappa_\tau \int f_{n_e} \, dl \, ,
\ee
with $\kappa_\tau = \sigma_T n_e^0$, 
and one sees that for an iso-thermal cluster, where $f_{T_e}(r) \equiv 1$,
one has $y \sim \tau T_e^0$.

For any cluster observation one receives information from an integral
along the line of sight. Since the radial structures of clusters are
non-trivial this implies that one may have to define averaged quantities.
For X-ray observations the averaging procedure is often simplified
to emission-weighted quantities, scaled by the emissivity
which for large tempe\-ratures 
roughly scales~\cite{sar86} as $\epsilon \sim n^2T_e^{1/2}$
\be
\langle {\cal O} \rangle_{X} = \frac{\int {\cal O} \epsilon \, 
dl}{\int \epsilon \, dl} \, .
\ee
For the SZ effect the relevant scale is not the emissivity, but instead
the local Compton parameter, $n_e(r) T_e(r)$, 
and therefore the average of ${\cal O}$ is
given by
\be
\langle {\cal O} \rangle_{y} 
= \frac{\int {\cal O} f_y \, dl}{\int f_y \, dl} \, ,
\label{eq:av}
\ee
for instance, the average SZ temperature for a cluster is
\be
\langle T_e \rangle_{y} = \frac{\int T_e f_y \, dl}{\int f_y \, dl} 
= \frac{\kappa_y \int T_e f_y \, dl}{y} \,.
\label{eq:avtemp}
\ee
We will hereafter only discuss Compton-averaged quantities, and hence 
omit the index on averages.

The SZ effect for an iso-thermal cluster is composed of several parts
\be
\frac{\Delta I}{I_0} = y \left( g(x) + T_e \, \delta(x,T_e) \right) 
- \beta_p \tau  \, a(x) \, ,
\label{eq:di1}
\ee
where $\beta_p = v_p/c$ is the bulk motion of the cluster referred to as
the peculiar velocity.
The spectral forms due to the thermal up-scattering, $g(x)$, 
and due to the overall cluster motion, $a(x)$, are well-known~\cite{sz,sz80},
and the spectral form of 
the relativistic corrections to the thermal up-scattering,
$\delta (x,T_e)$, is easily 
calculated~\cite{wri79,rep95,dol00,ito03,shi03}. To a very good 
approximation $\delta(x,T_e)$ can be taken as independent
of temperature (see e.g. Diego et al.~\shortcite{diego02}), 
that is, for a given frequency, $x$,
one can write $\delta(x,T_e) = b_1(x) + b_2(x) T_e + {\cal O}(T^2_e)$,
where $b_2$ and higher order terms are subdominant. 
For the present discussion we can ignore a contribution to
eq.~(\ref{eq:di1}) of the marginally detectable ultra-relativistic
electrons~\cite{ensslin1,ensslin2}, since our findings will not
depend on them.
For excellent reviews on the SZ effect see~\cite{birkinshaw,chr}.

For a cluster with a non-trivial temperature profile 
eq.~(\ref{eq:di1}) is written as
an integral along the line of sight
\be
\frac{\Delta I}{I_0} = \int \left[ 
\kappa_y f_y \left( g(x) + T_e^0 f_{T_e} \delta(x) 
\right)
- \beta_p \kappa_\tau f_{n_e} a(x) \right] \,dl \, ,
\label{eq:di}
\ee 
where the quantities $f_{n_e}(r)$ and $f_{T_e}(r)$ are the local
quantities. Now, using the definition in eq.~(\ref{eq:av}), this means 
\be
\frac{\Delta I}{I_0} = y \left[ g(x) + \langle T_e \rangle
\delta(x) - \beta_p \langle 1/T_e \rangle \, \left( 
m_ec^2/k\right) \, a(x) \right] \, .
\ee 
For an iso-thermal cluster, where $f_{T_e} \equiv 1$,
one has $1/\langle T_e \rangle =
\langle 1/T_e \rangle$,
but for a thermally non-trivial cluster there
are really 4 independent variables, $y, \beta_p, \langle T_e \rangle$
and $\langle 1/T_e \rangle$. Clearly, for such a cluster the X-ray
derived temperature, $T_X$, may differ significantly from 
$\langle T_e \rangle$ (as
well as, $1/T_X \neq \langle 1/T_e \rangle$), and hence the X-ray
derived temperature should only be used with great caution in the
study of SZ observations.

Most studies of SZ observations include up to 2 free 
parameters~\cite{laroque02,coma,battistelli}, 
for instance the Compton parameter and the peculiar velocity,
and only very few have attempted 
studies with 3 parameters~\cite{han02} with inclusion of the temperature.
We will here consider the effect of neglecting the non-trivial 
radial temperature profile, and show that it induces a systematic
error on the peculiar velocity of the order $10\%$.

\section{Cluster profiles}
\label{sec:cluster}
As a specific example we use a simple cluster model, where the
electron density profile follows a $\beta$-model
\be
n_e(r) = n_e^0 \, \left( 1 + \left( 
\frac{r}{r_c}\right)^2 \right)^{-\beta 3/2} \, ,
\label{eq:beta}
\ee
where $r_c$ is a characteristic radius. Such broken power-law
density profiles are
often used to fit observations~\cite{cavaliere,carollo,lewis}, and are possibly
even understood theoretically~\cite{hansenstadel,blois,rusz}, but may
need slight modifications for non-isothermal clusters~\cite{ettori2000}.
The  temperature profile may have the shape
\be
T_e(r) = T_e^0 \, \left( 1 + \frac{r}{r_c} \right) ^{-\alpha} \, ,
\label{eq:tempprof}
\ee
where we use results from recent numerically
simulated clusters~\cite{lin03} with
$\alpha = 0.56$ and $\beta = 0.61$. 
For cooling flow clusters the form is slightly more complicated,
but an inclusion of the central decrement is straight forward.
With these profiles one
finds the average temperature from eq.~(\ref{eq:avtemp}) to be
$\langle T_e \rangle =  7.08$ keV
when the central temperature is $10$ keV, and 
$\langle 1/T_e \rangle \langle T_e \rangle = 1.12$ which indicates
that one will find a peculiar velocity which is approximately $12\%$ 
systematically too large when using SZ observations to define the cluster
parameters. We will now test these findings numerically, by constructing
and analysing the true SZ signal from this cluster model.

With this cluster model, where we normalize the optical depth to $\tau =0.01$,
we construct the SZ signal along the line
of sight through the cluster centre. As a specific example we
take an observation with 4 observing frequencies at $90,150, 220$
and $270$ GHz, and assume a sensitivity of $1\mu K$ for each channel.
We choose $v_p = 1000$ km/sec.
For optimal observing frequencies~\cite{2003JCAP...05..007A,holder}
the statistical error-bars can be reduced slightly (represented by
$\Delta \chi^2 =1$ on figure 1), but the
systematic shift (about $12\%$) will remain the same.
The systematic shift is independent of the assumed sensitivity, 
which here is taken slightly optimistic (ACT expects a sensitivity
of $2\mu K$).

\begin{figure}
\begin{center}
\epsfxsize=8.5cm
\epsfysize=6.5cm
\epsffile{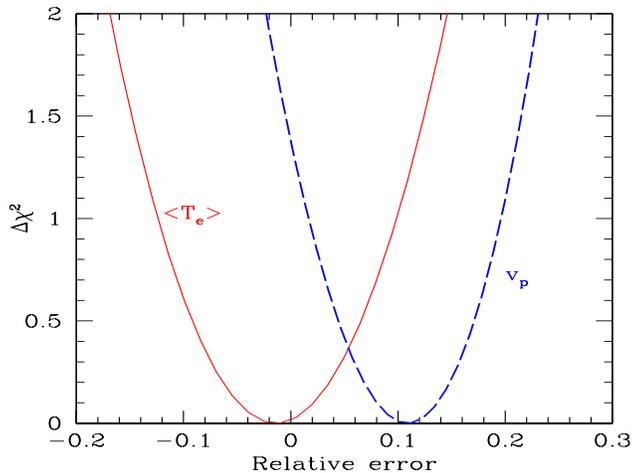}
\end{center}
\caption{The relative error on temperature, $\langle T_e \rangle$, and peculiar
velocity, $v_p$, through observation of a model SZ cluster with
non-trivial temperature profile (eq.~(\ref{eq:tempprof}) 
with $\alpha=0.56$) and
a $\beta$-profile for the electron density (eq.~\ref{eq:beta} 
with $\beta=0.61$).
The SZ-derived temperature (solid, red line) 
is very close to the Compton weighted
temperature, but the peculiar velocity (dashed, blue line) 
is systematically shifted
by $10-12\%$.}
\label{fig:1000}
\end{figure}

One can now treat this constructed SZ signal (and corresponding 
error-bars) as a real observation, and derive
the 3 cluster parameters, $y,T_e, v_p$ in the standard way. 
This is simply done by inputting
the constructed SZ signal into the publicly 
available~\footnote{{\tt http://krone.physik.unizh.ch/\~{}hansen/sz/} \\
This SZ parameter extraction code uses the Monte Carlo like method of
simulated annealing to derive the central SZ parameters and corresponding
error-bars.}
SZ parameter extraction code {\tt sasz}~\cite{Hansen:2003ff}. 
The result is shown in figure 1,
where we plot both peculiar velocity and temperature on the same figure.
The solid (red) line is the relative error on the temperature,
$(T_e-7.08)/7.08$, and the dashed (blue) line is the relative error 
on the peculiar velocity, $(v_p-1000)/1000$. As is clear, the average
temperature is indeed very near the expected $7$ keV, whereas the
peculiar velocity is shifted by $10-12\%$.

The fact that the SZ determined temperature indeed turns out to be
the ``correct'' Compton weighted temperature implies that
accurate prior temperature knowledge (e.g. from X-ray observations) would
not change the systematic shift of the peculiar velocity.
Having both accurate tempe\-rature and electron density profiles from
X-ray observations would allow one to calculate the systematic
shift (ignoring the complications from an unknown clumping). However,
peculiar velocities are most important when derived for a large sample of
clusters (e.g. from an all-sky SZ survey), but detailed X-ray maps are
observationally expensive and cannot be made for the large number of
clusters expected in future all-sky SZ surveys.

The minimal deviation from the derived central value of the
temperature arises because $\delta(x)$ in eq.~(\ref{eq:di}) is not
completely independent of the temperature, but this minor effect is
not important for our discussion. Furthermore, if one 
imposes a ``known'' temperature
of $T_e = T_X = 10$ keV (from X-ray observations)
then that would lead to a $\sim 40\%$ systematic shift of $v_p$, and even
a ``known'' emission weighted temperature of $T_X \approx 8$ keV
would give a $\sim 20\%$ systematic error on $v_p$.
To be very explicit, this means that if one knows the exact 
temperature and electron density profiles from X-ray observations,
and then uses the emission weighted temperature in the analysis
of SZ observations, then one would overestimate the peculiar
velocity with $20\%$ for this particular cluster.
If a given observation 
only puts bounds on the peculiar velocity, then the corresponding
error-bars are overestimated by the same amount.
This shows, as mentioned earlier, 
that X-ray observed temperatures should be use with
great caution in the study of SZ observations.
If one does know the exact temperature and density profiles, 
then clearly one can calculate $\langle 1/T_e \rangle$, with which
one can extract the peculiar velocity correctly.

We thus see that for this specific choice of temperature and electron
density profiles the systematic shift in the derived peculiar velocity
from SZ observations is $10 - 12 \%$. The question now arises:
which range in systematic shift is expected for real clusters?

The density profiles of many clusters are well described by a
$\beta$-model with $\beta \sim 0.67$, however, numerical 
simulations show a rather large scatter in the numerical value
for $\beta$. Fixing the temperature
profile to the one in eq.~(\ref{eq:tempprof}) with $\alpha=0.56$,
and letting $\beta$ vary from $\beta=0.75$ to $\beta=0.46$~\cite{lin03}
leads to systematic shifts in the peculiar velocity of $8\%$ to $20\%$.
One could instead fix the density slope to $\beta = 2/3$, and 
use a polytropic equation of state
\be
T_e = T_e^0 \, \left( \frac{n_e}{n_e^0} \right) ^{\gamma-1} \, ,
\label{eq:poly}
\ee
with $\gamma$ in the range $1$ to $5/3$. 
Even though the polytropic shape does not provide an excellent fit to
observations~\cite{degrandi02}, it is sufficient for estimating
the magnitude of the systematic error. Fitting observations with
eq.~(\ref{eq:poly}) gives
$\gamma = 1.46 \pm 0.06$ for non-cooling flow clusters, and
$\gamma = 1.20 \pm 0.06$ for cooling flow clusters~\cite{degrandi02}. 
With $\gamma = 1.46$ one gets a systematic shifts of the peculiar velocity
of $20\%$ (for $\gamma=1.20$ one gets $6\%$), 
and for the theoretically extreme case of an adiabatic gas with 
$\gamma = 5/3$ the error becomes about $30\%$. A simple fit to
the error in percent of the peculiar velocity 
as a function of $\gamma$, which is accurate for 
$\gamma>1.2$, gives: error (in $\%$) $=53(\gamma -1)-5$.

Cluster merging induces bulk motion within the cluster as large as
$500$ km/sec (see e.g. Sunyaev et al.~\shortcite{sunyaev}), and the 
corresponding
mis-estimate of the peculiar velocity is about $10\%$~\cite{tegmark},
or up to about $100$ km/sec~\cite{holder}.

\section{Onion peeling a SZ cluster}
\label{sec:onion}
We have seen above that the SZ determined peculiar velocity of
realistic galaxy clusters may be systematically wrong by $10-20\%$, and that
X-ray observations cannot be expected to solve this problem because
they are observationally expensive. The
natural question is then: what is the solution?

For X-ray observations the similar problem is solved by the technique
of {\it deprojection}. This heuristically corresponds to peeling
an onion. The outermost layer is observed and analysed. Then the
next layer is analysed while subtracting the signal from the 
outer layer. The error-bars will increase for the inner layers,
but for X-rays this turns out not to be disastrous. That is because the
contribution from the outer layers is significantly smaller than
the contribution from the inner layers, since the emission drops
fast with radius $\epsilon \sim T^{1/2} n^{2} \sim r^{\zeta_X}$ with
 $\zeta_X \sim -4.2$.

Similarly, for SZ observations the relevant quantity is the local
Compton parameter, which goes like $n_e T_e \sim r^{\zeta_y}$,
with $\zeta_y \sim -2.5$. Thus, the error-bars should remain under
control even when subtracting contribution to the intensity
from outer layers. Naturally, the fact that $2.5 < 4.2$ is the
reason why SZ observations are more suitable than X-ray observations
for measuring the outer cluster region.

An advantage for SZ deprojection 
is that the peculiar velocity is the same for the
entire cluster when we can ignore the local gas motion
(see however Holder~\shortcite{holder}).  That
means that while normally one would expect to have 3 SZ parameters
($y,T_e,v_p$) for each of the $N$ radial bins, giving $3N$ parameters,
then we really only have $2N+1$ free parameters. This is important
because the major parameter degeneracy for SZ observations is between
the temperature and peculiar velocity. Thus, the deprojection itself
may help reduce the error-bars on the derived parameters.
It is worth mentioning that spatially resolved SZ observations will
provide both temperature and electron density profiles, so combined
with accurate X-ray observations one can directly measure the radial
dependence of clumpiness, which is a non-trivial measure of the merger
history.

Such SZ deprojection of the observed intensity clearly demands that
the future SZ observations should both have good {\em spectral
coverage} and good {\em spatial resolution}. That is a non-trivial
observational challenge.

\section{Conclusion}
We have shown that due to the non-trivial temperature structure of
galaxy clusters, the SZ derived peculiar velo\-city will be
systematically shifted by $10-20\%$.  The Compton weighted electron
temperature is, however, derived accurately.  For future all-sky SZ
surveys one cannot rely on the observationally expensive X-ray
observations to remove this systematic error, but one should instead
reach for sufficient angular resolution to perform a deprojection in
the SZ spectra.

\section*{Acknowledgments}
It is a pleasure to thank 
Philippe Jetzer 
for discussions, and Elia  Battistelli for suggestions.
The author thanks the Tomalla foundation for financial support.

\label{lastpage}

\end{document}